\documentstyle[editedvolume]{crckapb} 

\begin{opening}
\title{DISTANCES FROM THE CORRELATION BETWEEN GALAXY LUMINOSITIES AND
       ROTATION RATES}

\author{R. Brent TULLY}
\institute{Institute for Astronomy, University of Hawaii\\
           2680 Woodlawn Drive, Honolulu, Hawaii 96822, USA}

\end{opening}

\runningtitle{LUMINOSITY--LINEWIDTH CORRELATIONS}

\begin{document}


\centerline{\bf Abstract}

\medskip\noindent{\it
A large luminosity--linewidth template sample is now available, improved
absorption corrections have been derived, and there are a statistically
significant number of galaxies
with well determined distances to supply the zero point.  A revised 
estimate of the Hubble Constant is 
${\rm H}_0 = 77 \pm 4$~km s$^{-1}$~Mpc$^{-1}$
where the error is the 95\% probable statistical error.  Systematic 
uncertainties are potentially twice as large.
}

\section{Introduction}

More massive galaxies have more stars than less massive galaxies and more
massive galaxies rotate faster.  This simple reasoning leads to the
expectation of a correlation between the light of the stars and a measure
of the rotation rate (Tully \& Fisher 1977) though the small scatter of the 
correlation is not trivially explained.  The measurement of the rotation rate,
say from the width of a global neutral hydrogen profile, is independent of
distance.  Hence, if the relationship is calibrated in terms of the
{\it intrinsic} luminosity dependence with linewidth then the modulus between
the apparent and absolute magnitude at a given linewidth gives a distance.

Nowadays there are more accurate methods for measuring the distance to an
individual galaxy.  However, the great virtue of the luminosity--linewidth
method is that it can be used to get distances for thousands of galaxies
all over the sky.  Roughly 40\% of galaxies with $M_B<-16^m$ are potential
targets.  Current techniques allow candidates to be accessed out to 
$V \sim 10,000$~km~s$^{-1}$.  Consequently, a determination can be made of
the Hubble Constant based on distance measurements dispersed on the sky and
in a regime with modest consequences due to peculiar velocities.

It is a good moment to review the results obtained with this method.  Lately
there has been a big input of high quality data.
Coupled with this happy situation is the dramatic improvement in the absolute
calibration as a result of the determination of distances by the cepheid
pulsation method with Hubble Space Telescope.  These two improvements address
what were the greatest deficiencies in the path to the Hubble Constant based
on the luminosity--linewidth method.  It will be shown during the ensuing
discussion that today there are sufficiently large numbers of both calibrators
and targets that the statistical accuracy of the method is very good.
Uncertainties are now dominated by potential systematic effects.

Here is an outline of the structure of the article.  First there will be a
discussion of the raw data: luminosities, axial ratios, and 
linewidths.  The information comes from many sources.  The observed parameters 
require adjustments for modifying effects like 
extinction and projection.  Then the methodology of the construction and
application of the luminosity--linewidth correlation will be described.
The potential problems of Malmquist bias must be confronted.  It will be
described how a template luminosity--linewidth correlation is constructed and
then transformed to an absolute magnitude scale.  Then the template relation 
is imposed upon data in clusters that for the most part are beyond the
Local Supercluster, at distances where peculiar velocities should be only a
modest fraction of expansion velocities.  That exercise results in distance
estimates for the clusters and a determination of the Hubble Constant.

\section{Data}

Three parameters must be measured: an apparent magnitude, a characterization
of the rotation rate, and an estimator of the inclination needed to 
compensate for projection effects.  Each of these components will be considered
in turn.  Then, there will be a discussion of the adjustments to be made to
get to the parameters that are used in the correlations.

\subsection{Luminosities}
Area photometry with optical and near-infrared imagers has come of age.  
Large format detectors on modest sized telescopes provide fields of view that
encompass the entire target galaxies.  The author has had an on-going program
of both optical and infrared photometry (Pierce \& Tully 1988, 1992;
Tully et al. 1996, 1998).  For the purposes of the present discussion, the
other important sources of luminosities are Mathewson,
Ford, \& Buchhorn (1992), Han (1992) and Giovanelli et al. (1997$b$).
The latter three sources provide $I$ band  magnitudes for galaxies in clusters
at intermediate to large distances.  The collaboration involving Pierce and
Tully is producing $B,R,I$ magnitudes for many nearby galaxies, including
the calibrators, and for galaxies in a couple of distant clusters that overlap
with the other sources.  In addition, data in the $K^{\prime}$ band is 
provided in Tully et al. (1996, 1998) for two of the clusters.

At the moment, there is a lot more data available at $I$ than at other bands
so most of the analysis presented in this paper will be based on this material.
There is interest in the other bands, though, because of the insidious effects
of obscuration.  There will be some comfort that there is proper compensation
for these effects if relative distances are the same at different passbands.
The $K^{\prime}$ material is of particular interest in this regard since
obscuration should be very small at 2~microns.

The issue of adjustments to magnitudes because of obscuration and spectral
shifting will be discussed in a later section.  The concern at this point is
the homogeneity of the raw magnitudes from various sources.  Different
authors measure magnitudes to slightly different isophotal levels then usually
extrapolate to total magnitudes:
Han extrapolates from $23.5^m$, Giovanelli et al. extrapolate from $\sim 24^m$,
Mathewson et al. extrapolate from $25.0^m$, Tully et al. (1996) extrapolate
from $25.5^m$, and Pierce \& Tully give a total magnitude to
sky at $\sim 26^m$.  The added light at the faintest levels is small for the 
high surface brightness galaxies that are relevant for the determination of
H$_0$.  Typical extrapolations from $25.5^m$ to infinity add $\sim0.02^m$ and
always less than $0.1^m$ for galaxies in the appropriate magnitude range
(Tully et al. 1996).  Magnitudes measurements are more vulnerable
to the detailed fitting of the sky level.  Variations at the level of 
$\sim 0.05^m$ can arise with systematic differences in sky fitting procedures.

Inter-comparisons between sources indicate that everybody is working on the
same system and that systematics are almost negligible.  Some offsets have
been reported, for example Giovanelli et al (1997$b$) adjust Mathewson et al.
data (1992) to match their own.  However, globally the data sets are consistent
with each other at a level of 2\% in effect on distances.  Object by object,
rms differences between any pair of observers is at or below $\pm 0.1^m$.
In the present analysis all sources are given
equal weight and luminosities are averaged if there are multiple
observations.  Overlap measurements reveal spurious results in a few
percent of cases.  If a difference between sources is big enough it
is usually evident which measurement is wrong.

\subsection{Inclinations}
Projection corrections are required to recover true disk rotation rates
and to compensate for differential obscuration.  Uncertainties in inclination
especially affect de-projected velocities as one
approaches face-on.  With rare exception, inclinations are derived from a
characteristic axial ratio of the main or outer body of a galaxy.  From
experience, it is found that such inclination measurements are reproduceable
at the level of $\pm 3^{\circ}$ rms.  However, errors are non-gaussian.
From the radial variations in axial ratios and from such independent 
considerations as inclination estimates from velocity fields it is suspected 
that errors of $\sim 10^{\circ}$ are not uncommon.  The $1/{\rm sin}i$
deprojection correction becomes grossly uncertain toward face-on.   
A sample cut-off at $i=45^{\circ}$ is invoked to avoid large errors.

The derivation of an inclination from an axial ratio requires an assumption
about the intrinsic thickness of the system.  The standard formulation
for the derivation is ${\rm cos}i = \sqrt{(q^2-q_0^2)/(1-q_0^2)}$ where
$q=b/a$ is the observed ratio of the minor to major axes and $q_0$ is the
intrinsic axial ratio.  The thinnest systems are spirals of type S$c$.
Earlier types have bulges and later types are puffed up.  For simplicity,
$q_0=0.20$ is often used.  A more elaborate specification of $q_0$ that
depends on type could be justified.  Giovanelli et al. (1997$b$) provide an
extreme example with their choice $q_0=0.13$ for type S$c$. 
A smaller $q_0$ value results in derived inclinations that are more face-on.  
Fortunately
the choice of $q_0$ has negligible effect on the measurement of distances 
as long as one is consistent.  The difference $q_0=0.13$ or 0.20
gives a difference in inclination, for an observed $q=0.20$, of
$81^{\circ}$ or $90^{\circ}$ respectively.  However the $1/{\rm sin}i$ 
difference on the corrected linewidth is only 1.2\%.  As one progresses
toward larger $q$ the difference in assigned inclination is reduced but the
$1/{\rm sin}i$ amplifier is growing.  The product of the two effects is a
roughly constant shift of 1.2\% in the corrected linewidth at all inclinations 
$i>45^{\circ}$.  If both calibrators and distance targets are handled
in the same manner there will be no effect on measured distances.

Difficulties with projection enter luminosities in the opposite regime, as 
galaxies are presented toward edge-on.  It has become popular to formulate
extinction corrections directly in terms of the observed $q$ value which
avoids a dependence on the parameter $q_0$.

Inter-comparisons between the sources of photometry used in this study
fail to reveal any systematic differences in $q$ measurements between authors,
though big individual differences are not uncommon.  Big differences raise 
flags that prompt special attention.
 
\subsection{Linewidths}
It has become popular to measure rotation parameters via both optical and radio
techniques.  The original radio methods are simpler but are constrained by
detector sensitivity to modest redshifts.  The methods that involve optical
spectra require more work but can be used to larger distances.  With care, 
the two techniques can be reconciled in a common characterization of the
projected rotation.  However that synthesis will not be attempted here.
There are plentiful and sufficiently distant observations of profiles 
in the 21~cm neutral hydrogen line for the purpose of determining
H$_0$.  The complexity of intermingling radio and optical data can be avoided.

Even restricted to the HI data, there is a bit of a mess.  For once in 
astronomy, resolution is not an unmitigated advantage.  In this case it is 
desired that the beam project onto an area larger than the galaxy in order to 
enclose most of the emission.  As a result, the data on nearby, large galaxies 
is handed down
from observations on old telescopes from the days of paper strips.  The
parameterizations are still quite `personalized'.  Worse than with magnitudes,
one has to be careful that one is using a consistent set of linewidth
information from the near field to far.  In this study, HI profile linewidths 
defined
at the level of 20\% of the peak flux are used (called $W_{20}$).  These 
linewidths are only 
adequately measured if the line signal-to-noise is greater than 7.  A
decent signal typically provides a measurement with an accuracy of better than
10~km s$^{-1}$.  The 20\% linewidths are then transformed into the parameter
$W_R$ defined by Tully \& Fouqu\'e (1985).  This parameter approximates
twice the maximum rotation velocity of a galaxy.

The other linewidth characterization in common use is the width at 50\% of
peak flux in each horn of the profile ($W_{50}$: Haynes et al. 1997) which is 
then adjusted to account for instrumental and thermal broadening (Giovanelli 
et al.
1997$b$).  The advantages and disadvantages of the alternative systems are 
technical and not important.  The key concern is that the information 
available over both north and south hemispheres and for both nearby large
galaxies and those distant and small be brought to a common system.
The current analysis draws on a large database of $W_{20}$ measurements
and a supplement of $W_{50}$ values in clusters well beyond the Local 
Supercluster.  The important new contributions in the $W_{50}$ system are
only partially within the public domain so it has not yet been possible to
make a detailed comparison based on the $\sim 10^3$ galaxies that must be 
mutually
observed in the two systems.  The present study relys on an inter-comparison
of only 66 galaxies in 3 clusters.  It is found that 
$<W_{20}-W_{50}>=25$~km~s$^{-1}$ with 12~km~s$^{-1}$ rms scatter and a
standard deviation of 2~km~s$^{-1}$.  There is no apparent trend with $W$
or between the three clusters.  While acceptable for a preliminary foray,
this inter-comparison of the $W_{20}$ and $W_{50}$ systems must, and can easily,
be improved upon.

\subsection{Extinction Corrections}
Along with the Hubble Space Telescope contribution to the zero-point 
calibration and the abundance of new material, the third significant
improvement of late has been in the compensation for extinction.
Giovanelli et al. (1995) made a convincing case for a strong luminosity
dependence in the obscuration properties of galaxies and Tully et al. (1998)
have further quantified the effect.  The latter work has profited from
the leverage provided by information in passbands from $B$ to $K^{\prime}$.
A giant galaxy can be dimmed by 75\% at $B$ if it is viewed edge-on rather than
face-on, although a dwarf galaxy of the luminosity of the Small Magellanic
Cloud statistically has not enough extinction at $B$ to measure.  At
$K^{\prime}$ the most luminous galaxy is dimmed by a maximum of 20\%.

\begin{figure}
\vspace{60mm}
\includegraphics{h0.fig1.eps}
\caption{
Dependence of the extinction amplitude parameter $\gamma_\lambda$ on
absolute magnitude.  Data are presented for $B,R,I$ bands in the three
separate panels.  The filled squares correspond to values of
$\gamma_\lambda$ derived from
deviations from mean color relations as a function of $b/a$.  The
filled triangles correspond to equivalent information derived from
deviations from luminosity--line profile width correlations as a
function of $b/a$.  The small circles in the $I$ panel are data taken
from Giovanelli et al. (1995).
The dashed straight line in the $I$ panel is a least
squares fit to the Giovanelli et al. data (errors in $\gamma$).  The
solid straight lines in this and the other panels are least squares
fits to the data by Tully et al. (1998).  The dotted straight
line in the $I$ panel gives equal weight to the two surces of data.
}\label{1}
\end{figure}

The extinction can be described by the expression 
$A_{\lambda}=\gamma_{\lambda}{\rm log}(a/b)$ where $a/b$ is the major to
minor axis ratio and $\lambda$ is the passband.  The correction is to
face-on orientation but does not account for the residual absorption in a
face-on system.
Figure~1 provides a plot from Tully et al. (1998) that shows the dependence
on luminosity of $\gamma_{\lambda}$ for $\lambda=B,R,I$.  Given this
strong luminosity dependence, there is a problem because absolute magnitudes
are not known a priori.  Absolute magnitudes are to be an output of the
distance estimation process so they
cannot also be an input.  Both Giovanelli et al. (1997$b$) and Tully et al. 
(1998)
recast the corrections for magnitudes so the dependency is on the
distance independent linewidth parameter.  This conversion is provided
through the luminosity--linewidth calibrators.  The formulations presented by
Tully et al. (1998) are:
\begin{equation}
\gamma_B = 1.57 + 2.75 ({\rm log} W_R^i - 2.5)
\end{equation}
\begin{equation}
\gamma_R = 1.15 + 1.88 ({\rm log} W_R^i - 2.5)
\end{equation}
\begin{equation}
\gamma_I = 0.92 + 1.63 ({\rm log} W_R^i - 2.5)
\end{equation}
\begin{equation}
\gamma_{K^{\prime}} = 0.22 + 0.40 ({\rm log} W_R^i - 2.5)
\end{equation}
There is a fortunate interplay that minimizes the effect of uncertain
inclination on $A_{\lambda}$.  If the inclination is taken too face-on because 
of an spuriously
large $b/a$ then $W_R^i$ is overestimated, which drives up $\gamma_{\lambda}$,
but is offset by a low ${\rm log}(a/b)$ in the product that gives 
$A_{\lambda}$.

The other corrections to be made are modest and non-controversial.
Absorption at $I$ due to obscuration in our own Galaxy is taken to be 41\% of 
the $B$ band value given by Burstein \& Heiles (1984).  There is a small 
`k-correction' of $1.27z$. 

\section{Methodology}
Over the years many people have used luminosity--linewidth relations to 
measure distances and there has been controversy.  An extreme view 
has been presented by Sandage (1994$b$).  According to him,
there can be large biases that distort distance measurements and limit the
usefulness of the procedure.  In this section there will be a description of
a way of conducting the analysis that results in unbiased distance estimates
and, hopefully, accurate results.  The reader interested in making a
comparison will find that the method to be described is {\bf not} the
method used by Sandage.

\subsection{Biases}
Malmquist (1920) discussed a bias that might create a problem with
measurements of distances to objects selected by apparent magnitude. Teerikorpi
(1984) and Willick (1994) have discussed the problem in the present context.
Schechter (1980) and Tully (1988$a$) have described a procedure that is
expected to {\it nullify} the bias.  That procedure will be summarized.

An example of when the bias arises is provided by considering the description 
of the luminosity--linewidth correlation given by the regression with errors
taken in magnitudes -- sometimes called the `direct' relation.  Use the
`direct' relation to determine distances to objects in the field.
By the construction of the regression, the brightest galaxies will tend to
lie above the correlation line.  Suppose one considers a group.  The 
brightest galaxies, drawn from above the mean correlation but assigned the
absolute magnitude of the mean correlation, will be given a {\it closer}
distance than is correct.  As fainter galaxies in the group are sampled,
they progressively sample the true distribution around the mean
correlation, so that the mean distances of the fainter galaxies are larger.
Kraan-Korteweg, Cameron, \& Tammann (1988) have shown that 
the measured mean
distance of a group increases as fainter objects are included.  For the same
reason, as one probes in the field to larger redshifts one samples 
progressively only the brightest galaxies, those that tend to be drawn from 
above the mean correlation. Hence one progressively assigns erroneously
low distances.  Low distances give a high H$_0$.

In an analysis made this way it is imperative that a
correction be made for the bias.  However, to make the correction it is
necessary to have detailed information on the form of the luminosity--linewidth
correlation and the nature of the scatter.  With adequate information, it
is possible to correct {\it statistically} for the bias, though the
trend of deviations with magnitude would persist in the individual 
measurements.  It is submitted that Sandage (1994$b$) provides an example
where the characteristics of the correlation and scatter are not 
understood and the corrections are erroneous.

Variations on the procedures that require bias corrections are pervasive
(eg, Willick et al. 1997).  For example, a maximum likelihood description
of the relationship (Giovanelli et al. 1997$b$) still retains the bias and 
requires corrections.  The corrections might be done properly.
However, these procedures require (1) that the
calibrators and targets have the same statistical properties, and (2)
detailed specification of the sources of scatter
and of properties of the luminosity function from which the sample is
drawn.
As an alternative, the method to be described {\it nulls} the bias 
rather than {\it corrects} for it.
Consequently, there is no
requirement to specify the sources of scatter or the properties of the sample. 
One is relying only on the
assumption that calibrators and targets have the same properties.

The magic description that nulls the bias is given by the regression with
errors in linewidth (Schechter 1980; Tully, 1988$a,b$) -- 
the `inverse' relation.  Two
qualitative comments might crystallize the merits of the 
procedure.  The first point to appreciate is that {\it the amplitude of the
bias depends on the assumed slope of the correlation.}  The flatter the 
dependence of magnitude with linewidth the greater the bias.  Conversely, 
if the slope is taken steep enough {\it the sign of the bias can be reversed.}
Hence it can be understood that there is a
slope that nulls the bias.  That slope is given by the regression on
linewidth if the sample is only limited in magnitude.
The second key point is made by a 
consideration of the regressions on the separate axes of a 
luminosity--linewidth plot.  Suppose one considers successively brighter
magnitude cuts on an intrinsic distribution.  As one progressively limits 
the magnitude range, the correlation coefficient of the fit will degrade.
Presented graphically, the correlations on the two axes will progressively
diverge as the fitting range is reduced.  Here is the critical point.
As the truncation is progressively advanced in magnitude {\it the slope 
of the correlation with errors in linewidths is always the same} but the slope
with errors in magnitudes is progressively splayed to shallower values.

Since the amplitude of the bias depends on the slope of the correlation, it
should be seen that an analysis based on the direct relation is on slippery
ground because the value of the slope depends on the magnitude limit of
the sample.  One needs a lot of information for an internally consistent
application.  The maximum likelihood approach raises the same qualitative 
concerns although, because it involves a slope intermediate between the
direct and inverse correlations, the quantitative problem is also
intermediate.

It has been pointed out by Willick (1994) that a bias can enter the inverse
correlation in practical applications.  The bias can be introduced because
the cutoff may not be strictly in magnitude.  For example, the sample might
be chosen at $B$ band but applied at a more redward band such as $I$.
A correlation between color and linewidth generates a slope to the magnitude
cutoff at a band other than $B$.  Or suppose the sample is selected by 
apparent diameter.  A correlation between surface brightness and linewidth
can again give a slope to the magnitude cutoff.  A slope in the magnitude
cutoff is equivalent to the introduction of a linewidth stricture.   Any 
restriction in linewidths brings the problem of bias over to the orthogonal
axis.  Two things can be said of this problem.  First it is a small effect,
down by a factor of five in amplitude in Willick's analysis.  Second
the problem is partially avoided by building the calibration out of only 
galaxies that satisfy a completion limit at the band to be considered; ie,
a stricter limit is taken than the one that provided the initial sample.

Most important: to achieve the correlation that nulls the bias one wants 
{\it a complete magnitude limited calibration
sample.}  In the population of the luminosity--linewidth diagram with the
calibration sample there should
not be any discrimination against candidates in any particular part of the
diagram above the magnitude limit.  Selection based on 
inclination is inevitable but that restriction should be distributed across 
the diagram.
Other potential restrictions must be considered in a similar light.

The good news is that, with due care to the calibration, then the 
method
can be applied to give unbiased distances to individual galaxies in the field
as long as the inclusion of those galaxies is not restricted in
linewidth.  In other words, there will {\it not} be a correlation
between luminosity and distance within a group as found be Kraan-Korteweg et 
al. (1988) nor a correlation between H$_0$ and redshift as found by Sandage 
(1994$a$).
The method will break down if the target galaxy is a dwarf
intrinsically fainter than the limit of the calibration.  
The latter issue is only a concern in our immediate
neighborhood, not for the H$_0$ problem.

\subsection{The Template Relation at $I$ Band}
The creation of the template relation is a critical step.  In the section on
biases it has been described how important it is to have a sample that only
suffers magnitude constraints.  Often the calibration
relationship is formed 
out of the ensemble of a field sample (Willick et al. 1996) but 
the constraints on such samples are usually ambiguous.  Also, the calibration
relationship is inevitably broadened and distorted by non-Hubble expansion 
motions.

Cluster samples have evident advantages.  It is possible to be complete 
to a magnitude limit and it can be assumed that the galaxies are 
all at the same {\it relative} distance.  The biggest concern with cluster 
samples is whether there are intrinsic differences between galaxies in a 
cluster environment and those that are more isolated.  An operational
disadvantage of cluster samples is that an individual cluster does not
provide enough systems to provide good statistics.  These two disadvantages
can be addressed simultaneously by building a template relation out of 
several cluster samples.  The `clusters' can have a sufficient range in their
properties that one can begin to evaluate the issue of environmental
dependence.  The combination of several cluster samples takes care of the
problem of poor statistics.

This study uses samples drawn from five clusters with reasonable
completion characteristics.  There is best control with the nearby
Ursa Major and Fornax clusters.  The completeness limits in Ursa Major are 
discussed by Tully et al. (1996) and in Fornax by Bureau, Mould, and
Staveley-Smith (1996).  After corrections for obscuration, and translation
to $I$ magnitudes, the completion limit for both clusters is $I=13.4^m$.
There are 38 galaxies in Ursa Major with type S$a$ or later and 
$i\ge 45^{\circ}$ above this limit.  There are 16 galaxies in Fornax 
satisfying these constraints.  It was appreciated in advance
that Ursa Major and Fornax are at similar distances.  Hence the 
apparent magnitude limits conform to about the same absolute magnitude 
limits.  Fornax is indicated by these data to be $0.10^m$ closer.

Already a diverse environmental range has been explored between the Ursa Major 
and Fornax cases.  Tully at al. (1996) have labored the point that the 
Ursa Major Cluster environment is more similar to that of low density spiral
groups than to what is generally considered a cluster.  The structure must be 
dynamically young.  By contrast, Fornax has a dense core of early type
systems, evidence of a dynamically evolved structure.  Granted, the spirals
in the Fornax sample are more widely distributed than the central core and
may represent recent arrivals.

The next component to be added to the template is drawn from the filament
that passes through what has been called the Pisces Cluster.  Aaronson et al.
(1986) and Han \& Mould (1992) have included the region in their distance 
studies but Sakai, Giovanelli, \& Wegner (1994) have shown that one is dealing
with an extended structure with separate sub-condensations.  It is unlikely
that the region as a whole is collapsed.  Indeed, what will be considered here 
is a length of $\sim 20^{\circ}$ along the Pisces filament, which  corresponds
to an end-to-end distance of $\sim 20$~Mpc.  The mean redshift is constant to
$\sim 4\%$ along the filament though individual redshifts scatter over a 
range of $\pm 20\%$ relative to the mean.  It can
be asked if the full length of the filament is at a common distance or if
variations in distance can be identified.  A luminosity--linewidth correlation 
is constructed for the ensemble, then inter-compared by parts to 
determine if components deviate significantly from the mean.  
There is not the slightest hint of deviations from the mean.  Six 
sub-components along the $20^{\circ}$ filament have consistent distances
within a few percent.  To within measurement errors, the filament is tangent 
to the plane of the sky in both real space and velocity space.

Given this circumstance, all the galaxies with $3700<V_{cmb}<5800$~km~s$^{-1}$
along the $20^{\circ}$ segment of the Pisces filament 
$00^h44^m<\alpha<02^h13^m$
will be taken to be at the same distance.  Failures of this assumption will 
act to increase the scatter of the luminosity--linewidth relationship but the 
scatter  is found to be only $0.31^m$, as small as for any sub-component
of the template.  This scatter is with 46 galaxies, after rejection of one 
object that deviates by $\sim 4\sigma$.
There is reasonable completion brighter than $I=13.8^m$ which
is taken as the magnitude limit for the present sample.  The Pisces filament 
data is added to the Ursa Major/Fornax template by (1) calculating the
offset from the slope of the 2 cluster template, (2) redetermining a new 
slope now with 3 clusters, (3) iterating the distance offset with the new
slope, and (4) iterating the new 3 cluster template slope.  The distance
shift at step 3 is of order $1\%$ and the slope shift at step 4 is $\sim 1\%$.

\begin{figure}
\vspace{120mm}
\includegraphics{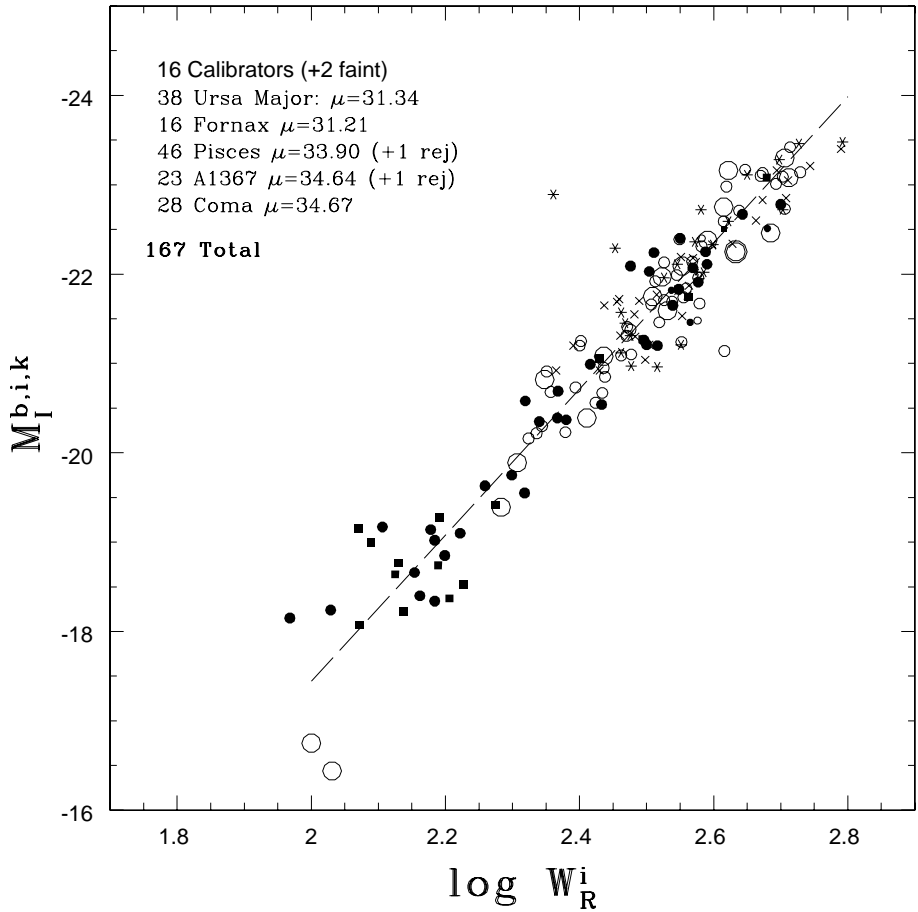}
\caption{
Template luminosity--linewidth relation at $I$ constructed with 151 galaxies
in five clusters, translated in zero-point to get a best fit with 16 
calibrators with accurate independent distances.  The slope is given by the
regression with errors in linewidths to the 151 cluster galaxies.  The
separate components of the plot are more easily seen in the figures that 
follow.
}\label{2}
\end{figure}

\begin{figure}
\vspace{67mm}
\includegraphics{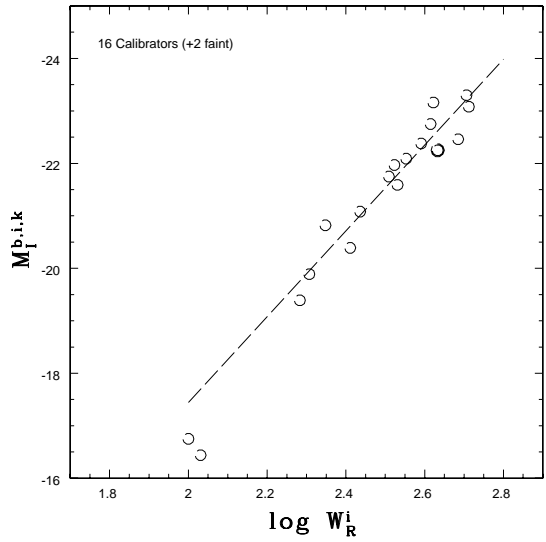}
\caption{
Luminosity--linewidth relation for distance calibrators.
}\label{3}
\end{figure}

\begin{figure}
\vspace{67mm}
\includegraphics{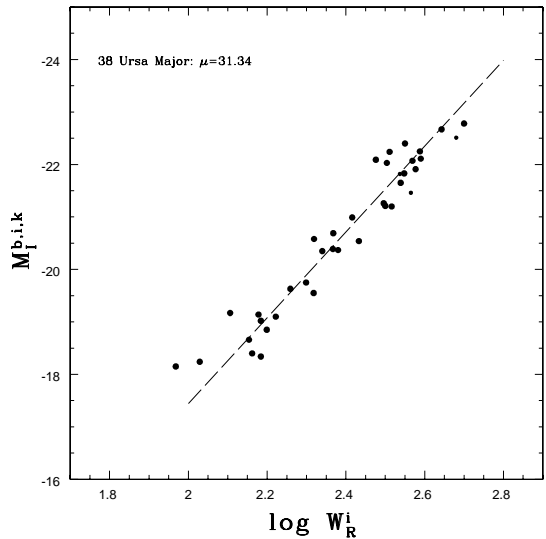}
\caption{
Luminosity--linewidth relation for Ursa Major Cluster.
}\label{4}
\end{figure}

\begin{figure}
\vspace{67mm}
\includegraphics{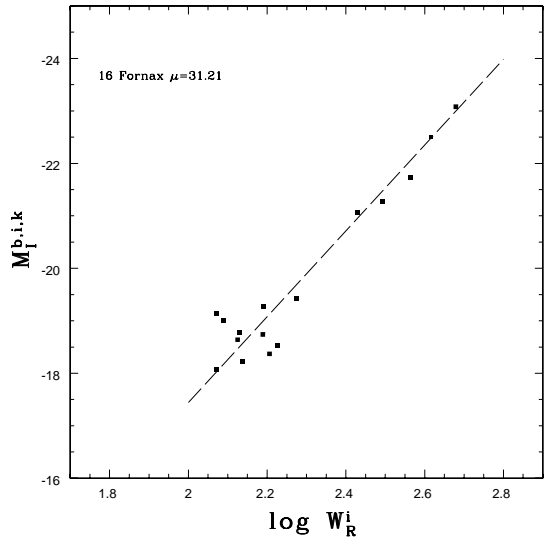}
\caption{
Luminosity--linewidth relation for Fornax Cluster.
}\label{5}
\end{figure}

\begin{figure}
\vspace{67mm}
\includegraphics{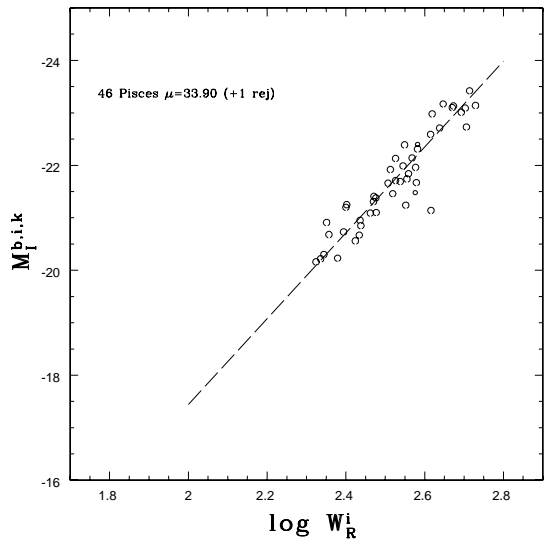}
\caption{
Luminosity--linewidth relation for Pisces filament.
}\label{6}
\end{figure}

\begin{figure}
\vspace{67mm}
\includegraphics{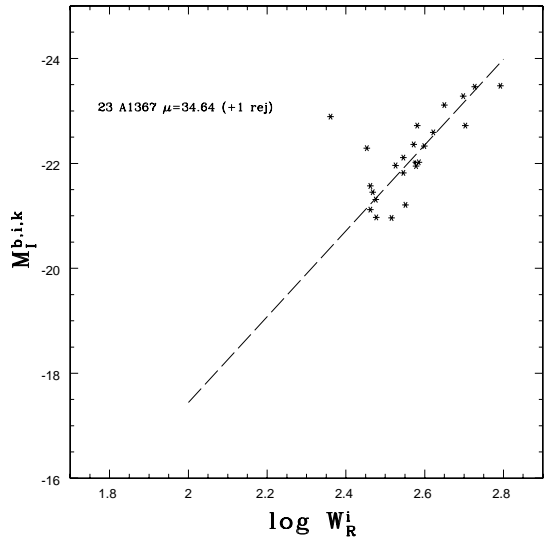}
\caption{
Luminosity--linewidth relation for Abell 1367 Cluster.
}\label{7}
\end{figure}

\begin{figure}
\vspace{67mm}
\includegraphics{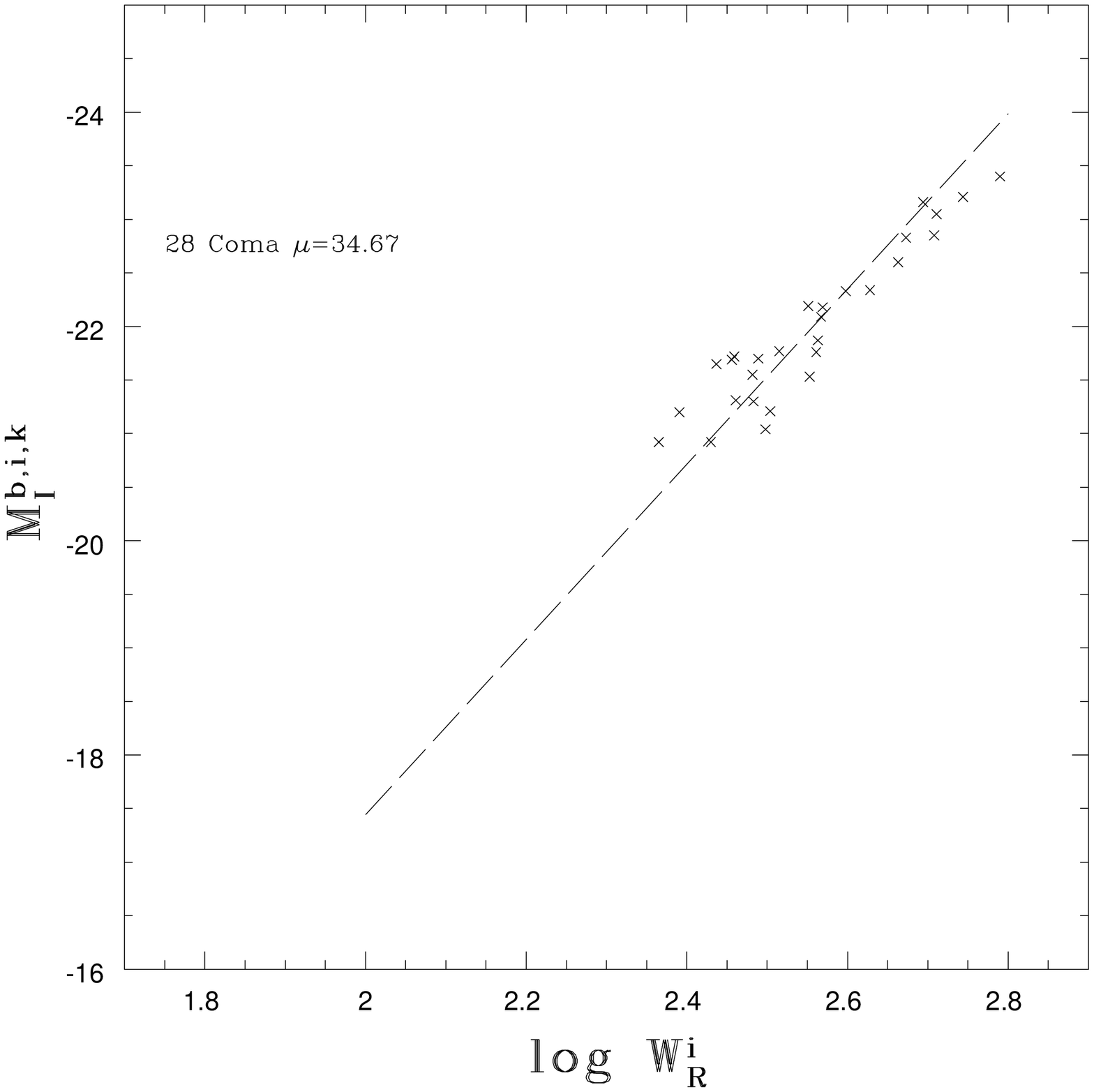}
\caption{
Luminosity--linewidth relation for Coma Cluster.
}\label{8}
\end{figure}

The final step in the development of the template is the addition of the
Coma and Abell~1367 clusters.  These clusters are at the same distance
to within a few percent so they are treated together until the final
iteration, at which point they are considered separately against the mean 
relation.
Only galaxies within $4.3^{\circ}$ of the cluster centers are accepted
and the velocity constraints described by Giovanelli et al. (1997$b$) are 
accepted.  As with Pisces, there is substantial but not full completion
to $I=13.8^m$.  
Iterations like those described with the Pisces filament 
converge to provide the five cluster template.  There could have
been a problem if there is curvature in the template, as might be
indicated if, say, the slope flattened for samples with more
luminous cutoffs (more distant clusters).  However there is no suggestion of
such a flattening.  The Coma sample provides 28 galaxies.  A1367 adds 23,
after one $5\sigma$ rejection.

In total, there are 151 galaxies in the 5 cluster template.  There are three
distinct absolute magnitude cutoffs (UMa/Fornax; Pisces; Coma/A1367) but,
to the degree that the slopes are indeed constant between components, the
calibration slope does not depend on the galaxy luminosity function.  If there
was evidence of a slope change, a slightly more complicated analysis
involving a non-linear relationship would have been necessary.

\subsection{Absolute Calibration at $I$ Band}
Curently there are 17 galaxies with distances determined through observations 
of cepheid variable stars, mostly from observations with the Hubble Space
Telescope (Freedman et al. 1997, Sandage et al. 1996, Tanvir et al. 1995).  
Two of the systems, NGC~2366 and NGC~3109, 
are fainter than the template cutoff so will be ignored.  One more galaxy
is added as a calibrator, NGC~4258, which has a distance from the geometry
inferred for the circum-nuclear masers (Miyoshi et al. 1995).  
Hence 16 calibrators are used.

It would be improper to do a regression on the calibrator relationship because
in no way do they provide a complete sample.  It can only be assumed that the
calibrators are drawn from a similar distribution as the template objects, 
perhaps with magnitude as a selection criterion but not linewidth.  
Effectively, each of the 16 calibrators provides a separate zero-point offset.
The least-squares average provides the optimum fit.  The final result is shown 
in Figure~2 where the $I$ band luminosity--linewidth relation is shown for 
the 16 calibrators and the 151 cluster template galaxies shifted to the 
absolute magnitude scale of the calibrators (the 2 rejected galaxies are
also plotted).  Figures~3-8 present the same material but separated to 
distinguish the fits to the calibrators and the individual clusters.

\subsection{The $B,R$ and $K^{\prime}$ Relations}

\begin{figure}
\vspace{180mm}
\includegraphics{h0.fig9.eps}
\caption{
Luminosity--linewidth relations in $B,R,I,K^{\prime}$ after corrections for
inclination.  {\it Filled circles:} Ursa Major; {\it open circles:}
Pisces.  Straight lines are regressions with errors in linewidths.
}\label{9}
\end{figure}

Less complete information is available at other bands than $I$.  However,
inter-comparisons are valuable because of the potential problem with
obscuration.  Information is available at $B$ and $R$ for the calibrators,
all the galaxies in the Ursa Major sample, most of those in Coma, and for most
in the part of the Pisces region at $00^h49^m<\alpha<01^h32^m$ (Pierce \& 
Tully 1998).
Material is available at $K^{\prime}$ for the same Ursa Major and Pisces 
galaxies (Tully et al. 1996, 1998).  The luminosity--linewidth relations
are illustrated in Figure~9.

The magnitude scatter is essentially the same at $R,I,K^{\prime}$ and 
$\sim 20\%$ worse at $B$.  Obscuration corrections diminish toward the
infrared until they are tiny at $K^{\prime}$.  However, sky background 
contamination causes degradation toward the infrared from the favorable 
situation at $R$ to the poor situation at $K^{\prime}$ where one looses
almost 2 scalelengths to the sky compared with an $R$ exposure of the same 
duration.
The correlations are seen to steepen toward the infrared.  However, this 
steepening is less extreme than had been seen in the past because of the 
strong luminosity dependence of the reddening corrections that are now applied.
The biggest corrections are made to the most luminous galaxies in the bluest
bands.  Hence the corrected relations at shorter wavelengths are steepened
toward the slopes of the almost-reddening-free infrared relations.  As shown
in Tully et al. (1998), only a weak color dependency on luminosity remains 
after reddening is taken into account.  Slopes at $B,R,I,K^{\prime}$
are -7.8, -8.0, -8.2, and -8.7, respectively, with the correlation against
the same linewidth information.  These slopes are based on the linewidth
regression which is appropriate for bias-free distance determinations but not
the slopes that one wants to give a physical interpretation.  The true nature
of the correlation is characterized better by a maximum likelihood fit.  An
approximation to that is a double regression, which at $I$ gives the slope 
-7.9.  These fits
indicate infrared convergence toward $L \propto W^n$ where $n=3.4 \pm 0.1$.

\section{Summary of Results}

The template relation with the zero-point given by 16 galaxies
can be used to determine distances to any other galaxy, with the proviso
that the proceedure must fail if the target is intrinsically less luminous
than $M_I^{b,i,k}=-18^m$, the faintness limit of the calibration.  If the
problem is to measure H$_0$, this faint limit is of no concern because
targets of interest are beyond the Local Supercluster where peculiar 
velocities are expected to be a small fraction of expansion velocities.

It would be possible to apply the calibration to measure distances to 
hundreds of galaxies.  For the moment, with the interest of maintaining
as homogeneous a set of measurements as possible, the H$_0$ determination
will be based on the 5 clusters that went into the template plus 7 other
clusters each with of order a dozen measures.  The results are presented
in Table~1 and Figure~10.  The table provides (col.~2) the number of 
measures in the cluster, (col.~3) the rms scatter about the template
relation, (col.~4/5) the distance modulus/distance of the cluster, 
(col.~6) the velocity of the cluster in the CMB frame as given by 
Giovanelli et al. (1997$b$), and (col.~7) the measure of H$_0$ from the 
cluster.
The velocity given to the Pisces filament is the average of the values for
the three main sub-condensations.

\begin{table}[htb]
\begin{center}
\caption{Five Template Clusters and Seven More}
\begin{tabular}{lrccccc}
\hline
Cluster & No. & RMS & Modulus & Distance & $V_{cmb}$ & H$_0$ \\
        &    & (mag) & (mag)  &  (Mpc)   & (km/s)  & (km/s/Mpc) \\
\hline
Fornax          & 16 & 0.50 & 31.21 & 17.5 &  1321  & 76\\
Ursa Major      & 38 & 0.41 & 31.34 & 18.5 &  1101  & 59\\
Pisces Filament & 46 & 0.31 & 33.90 & 60.3 & (4779) & 79\\
Abell 1367      & 23 & 0.41 & 34.64 & 84.8 &  6735  & 79\\
Coma            & 28 & 0.33 & 34.67 & 85.9 &  7185  & 84\\
\hline
Antlia          & 11 & 0.27 & 32.78 & 35.9 &  3120  & 87\\
Centaurus 30    & 13 & 0.52 & 32.94 & 38.9 &  3322  & 86\\
Pegasus         & 12 & 0.37 & 33.36 & 46.9 &  3519  & 75\\
Hydra I         & 11 & 0.35 & 33.84 & 58.6 &  4075  & 70\\
Cancer          & 16 & 0.34 & 33.89 & 60.0 &  4939  & 83\\
Abell 400       &  9 & 0.24 & 34.91 & 96.1 &  6934  & 72\\
Abell 2634      & 16 & 0.32 & 35.23 & 111.0 & 7776  & 70\\
\hline
{\bf Weighted average} & &  &       &      &        & {\bf 77}\\    
\hline
\end{tabular}
\end{center}
\end{table}

\begin{figure}
\vspace{120mm}
\includegraphics{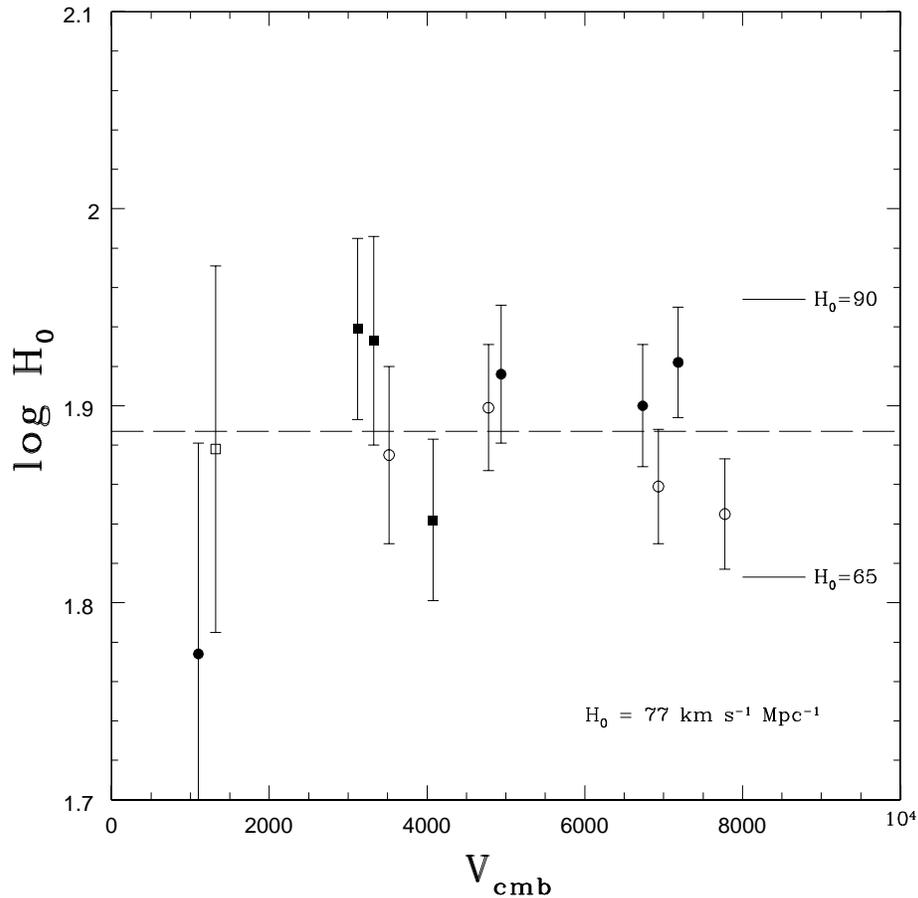}
\caption{
Individual estimates of H$_0$ as a function of systemic velocity.
Errors are a convolution of the statistical errors in distance and 
an uncertainty of 300~km s$^{-1}$ in velocities.  Symbols vary with location
on the sky: {\it filled circles:} north celestial and north galactic;
{\it open circles:} north celestial and south galactic; 
{\it filled squares:} south celestial and north galactic; 
{\it open squares:} south celestial and south galactic. 
}\label{10}
\end{figure}

The error bars in Fig.~10 contain distance and velocity components.
The errors associated with distance depend directly on the rms dispersion
in a cluster and inversely with the square root of the number of galaxies
in the cluster sample.  The error associated with velocity streaming is taken
to be 300~km~s$^{-1}$.  The velocity component to the error is totally dominant
inside 2000~km~s$^{-1}$.  The statistical errors in distance become the
dominant factor beyond $\sim 6000$~km~s$^{-1}$.  The symbols in Fig.~10
differ for different regions of the sky.  There is a hint of systematics:
for example the filled circles lie above the open circles.  For the present
purposes, the best estimate of H$_0$ is derived by taking an average of
log~H$_0$ values with weights proportional to the inverse square of the
error bars that are plotted.  The result is  
H$_0 = 77\pm 4$~km~s$^{-1}$~Mpc$^{-1}$.  The error is the 95\% probability
statistical uncertainty.  It is small because it is based on a
template of 151 galaxies, a zero-point fixed by 16 galaxies, and application
to 12 clusters distributed around the sky and out to 8,000~km~s$^{-1}$.

This result is somewhat higher than `interm' value of H$_0 \simeq 73$ 
reported by the HST Key Project team (Mould et al. 1997) or the value of
H$_0 = 69\pm 5$ found by Giovanelli et al. 1997$a$ based on similar 
applications of luminosity--linewidth correlations.  The latter value is
outside the bound of the statistical error found in this paper.  It must be
attributed to a small systematic difference that remains to be identified.

The value of H$_0 = 77$ found here is about 13\% lower than the value found
by the same author and method in the past.  There has been a 17\% decrease 
due to
the revision of the luminosity--linewidth zero-point as a consequence of
the increase 
from 4 to 16 in the number of calibrator galaxies with 
distances determined through the cepheid period--luminosity relation.
There has been an 8\% increase as a consequence of the revised reddening 
corrections now being applied.  A 5\% decrease has come about with the
introduction of material on more clusters around the sky at distances
well beyond the Local Supercluster.  These changes are a sobering illustration
of random and systematic errors.  The shift associated with the improved
cepheid calibration is comparable to the rms dispersion of the 
luminosity--linewidth relations at $R,I,K^{\prime}$ bands.  Either by
statistical fluke or an unknown systematic, the original 4 calibrators 
are drawn from the faint side of the intrinsic correlation.

In conclusion, there has been a decrease of $\sim$~one standard deviation
from the value of H$_0$ measured previously by this author with the 
luminosity--linewidth method, to H$_0 = 77 \pm 4$ km~s$^{-1}$~Mpc$^{-1}$.
The greatest perceived problems of old have been addressed: there are now
many more zero-point calibrators, the template is much more extensive and
complete, reddening corrections are under better control, and the method is
applied to more targets distributed around the sky.  Formal statistical errors
are cut in half.  Uncertainties are now dominated by potential, unidentified
systematics.

\section{References}

\noindent
Aaronson, M., Bothun, G., Mould, J.R, Huchra, J.P, Schommer, R.A., \&

Cornell, M.E.  1986, {\it Astrophy. J.}, {\bf 302}, 536.

\noindent 
Bureau, M., Mould, J.R., \& Staveley-Smith, L.  1996, {\it Astrophy. J.}, 
{\bf 463}, 60.

\noindent
Burstein, D., \& Heiles, C.  1984, {\it Astrophy. J. Suppl.}, {\bf 54}, 33.

\noindent
Freedman, W.L., Mould, J.R., Kennicutt, R.C. Jr., \& Madore, B.F.  1997,

in {\it IAU Symp. 183: Cosmological Parameters and the Evolution of the

Universe}, Kyoto, Japan.

\noindent
Giovanelli, R., Haynes, M.P., da Costa, L.N., Freudling, W. Salzer, J.J., \&

Wegner, G. 1997$a$, {\it Astrophy. J.}, {\bf 477}, L1.

\noindent
Giovanelli, R., Haynes, M.P., Herter, T., Vogt, N.P., Wegner, G., 
Salzer, 

J.J., da Costa, L.N., \& Freudling, W. 1997$b$, {\it Astron. J.}, {\bf 113}, 
22.

\noindent
Giovanelli, R., Haynes, M.P., Salzer, J.J., Wegner, G., da Costa, L.N., \& 

Freudling, W. 1995, {\it Astron. J.}, {\bf 110}, 1059.

\noindent
Han, M.S. 1992, {\it Astrophy. J. Suppl.}, {\bf 81}, 35.

\noindent
Han, M.S. \& Mould, J.R. 1992, {\it Astrophy. J.}, {\bf 396}, 453.

\noindent
Haynes, M.P., Giovanelli, R., Herter, T., Vogt, N.P., Freudling, W., Maia, 

M.A.G., Salzer, J.J., \& Wegner, G. 1997, {\it Astron. J.}, {\bf 113}, 1197.

\noindent
Kraan-Korteweg, R.C., Cameron, L.M., \& Tammann, G.A. 1988, {\it Astrophy. 

J.},
{\bf 331}, 620.

\noindent
Malmquist, K.G.  1920, Medd. Lunds Ast. Obs. Series II, no. 22.

\noindent
Mathewson, D.S, Ford, V.L, \& Buchhorn, M.  1992, {\it Astrophy. J. Suppl.}, 

{\bf 81}, 413.

\noindent
Miyoshi, M., Moran, J.M., Herrnstein, J.R., Greenhill, L., Nakai, N., 

Diamond, P., \& Makoto, I. 1995, {\it Nature}, {\bf 373}, 127.

\noindent
Mould, J.R., Sakai, S., Hughes, S., \& Han M.S.  1997, in {\it The 
Extragalactic 

Distance Scale: STScI Symp.~10}, M. Livio, M. Donahue, N. Panagia, 

(Cambridge U. Press), p. 158. 

\noindent
Pierce, M.J., \& Tully, R.B.  1988, {\it Astrophy. J.}, {\bf 330}, 579.

\noindent
Pierce, M.J., \& Tully, R.B.  1992, , {\it Astrophy. J.}, {\bf 387}, 47.

\noindent
Pierce, M.J., \& Tully, R.B.  1998, in preparation.

\noindent
Sakai, S., Giovanelli, R., \& Wegner, G.  1994, {\it Astrophy. J.}, 
{\bf 108}, 33.

\noindent
Sandage, A. 1994$a$, {\it Astrophy. J.}, {\bf 430}, 1.

\noindent
Sandage, A. 1994$b$, {\it Astrophy. J.}, {\bf 430}, 13.

\noindent
Sandage, A., Saha, A., Tammann, G.A., Labhardt, L., Panagia, N., \&

Macchetto, F.D.  1996, {\it Astrophy. J.}, {\bf 460}, L15.

\noindent
Schechter, P.L.  1980, {\it Astron. J.}, {\bf 85}, 801.

\noindent
Tanvir, N.R., Shanks, T., Ferguson, H.C., \& Robinson, D.R.T.  1995,

{\it Nature}, {\bf 377}, 27.

\noindent
Teerikorpi, P. 1984, {\it Astron. Astrophys.}, {\bf 141}, 407.

\noindent
Tully, R.B. 1988$a$, {\it Nature}, {\bf 334}, 209.

\noindent
Tully, R.B. 1988$b$, in {\it The Extragalactic Distance Scale}, Eds. S. van den

Bergh and P.J. Pritchet, ASP Conf. Ser. {\bf 4}: 318-328.

\noindent
Tully, R.B., \& Fisher, J.R.  1977, {\it Astron. Astrophys.}, {\bf 54}, 661.

\noindent
Tully, R.B., \& Fouqu\'e, P. 1985, {\it Astrophy. J. Suppl.}, {\bf 58}, 67.

\noindent
Tully, R.B., Pierce, M.J., Huang, J.S., Saunders, W., Verheijen, M.A.W.,

\& Witchalls, P.L. 1998, {\it Astron. J.}, {\bf 115}, (June).

\noindent
Tully, R.B., Verheijen, M.A.W., Pierce, M.J., Huang, J.S., \& Wainscoat,
 
R.J.  1996, {\it Astron. J.}, {\bf 112}, 2471.

\noindent
Willick, J.A.  1994, {\it Astrophy. J. Suppl.}, {\bf 92}, 1.

\noindent
Willick, J.A, Courteau, S., Faber, S.M., Burstein, D., Dekel, A., \& Kolatt, 

T.  1996, {\it Astrophy. J.}, {\bf 457}, 460.

\noindent
Willick, J.A, Courteau, S., Faber, S.M., Burstein, D., Dekel, A., \& Strauss, 

M.A. 1997, {\it Astrophy. J. Suppl.}, {\bf 109}, 333.

\end{document}